%% file: sample-sigconf.tex
\documentclass[sigconf]{acmart}
\usepackage[ruled,linesnumbered]{algorithm2e}
\usepackage{booktabs} 
\usepackage{multirow}
\AtBeginDocument{%
  \providecommand\BibTeX{{%
    \normalfont B\kern-0.5em{\scshape i\kern-0.25em b}\kern-0.8em\TeX}}}

\copyrightyear{2021}
\acmYear{2021}
\setcopyright{rightsretained}
\acmConference[GECCO '21]{2021 Genetic and Evolutionary Computation Conference}{July 10--14, 2021}{Lille, France}
\acmBooktitle{2021 Genetic and Evolutionary Computation Conference (GECCO '21), July 10--14, 2021, Lille, France}
\acmDOI{10.1145/3449639.3459370}
\acmISBN{978-1-4503-8350-9/21/07}

\begin{document}

\title[A Systematic Comparison Study on Hyperparameter Optimisation of Graph Neural Networks]{A Systematic Comparison Study on Hyperparameter Optimisation of Graph Neural Networks for Molecular Property Prediction}

\author{Yingfang Yuan}
\affiliation{%
  \institution{{School of Mathematical and Computer Sciences \\Heriot-Watt University}}
  \streetaddress{First Gait}
  \city{Edinburgh}
  \country{United Kingdom}
  \postcode{EH14 4AS}}
\email{yyy2@hw.ac.uk}

\author{Wenjun Wang}
\affiliation{%
  \institution{School of Mathematical and Computer Sciences \\
  Heriot-Watt University}
  \streetaddress{First Gait}
  \city{Edinburgh}
  \country{United Kingdom}
  \postcode{EH14 4AS}}
\email{wenjun.wang@hw.ac.uk}

\author{Wei Pang}
\authornote{Corresponding author}
\affiliation{%
  \institution{School of Mathematical and Computer Sciences \\
  Heriot-Watt University}
  \city{Edinburgh}
  \country{United Kingdom}
  \postcode{EH14 4AS}}
\email{w.pang@hw.ac.uk}

\renewcommand{\shortauthors}{Yingfang, et al.}

\begin{abstract}
Graph neural networks (GNNs) have been proposed for a wide range of graph-related learning tasks. In particular, in recent years, an increasing number of GNN systems were applied to predict molecular properties. However, a direct impediment is to select appropriate hyperparameters to achieve satisfactory performance with lower computational cost. Meanwhile, many molecular datasets are far smaller than many other datasets in typical deep learning applications. Most hyperparameter optimization (HPO) methods have not been explored in terms of their efficiencies on such small datasets in the molecular domain. In this paper, we conducted a theoretical analysis of common and specific features for two state-of-the-art and popular algorithms for HPO: TPE and CMA-ES, and we compared them with random search (RS), which is used as a baseline. Experimental studies are carried out on several benchmarks in MoleculeNet, from different perspectives to investigate the impact of RS, TPE, and CMA-ES on HPO of GNNs for molecular property prediction. In our experiments, we concluded that RS, TPE, and CMA-ES have their individual advantages in tackling different specific molecular problems. Finally, we believe our work will motivate further research on GNN as applied to molecular machine learning problems in chemistry and materials sciences.
\end{abstract}



\maketitle
\input{main}

\bibliographystyle{ACM-Reference-Format}
\bibliography{mybib} 
\end{document}

%% file: main.tex
\section{Introduction}
Graph neural networks (GNNs) are efficient approaches for learning the representations of structured graph data (e.g., molecules, citation networks) \cite{xu2018powerful}. In recent years, several types of GNNs have been proposed for predicting molecular properties, and they have achieved excellent results~\cite{duvenaud2015convolutional, kearnes2016molecular, wu2018moleculenet, gilmer2017neural, lamb2020bayesian, hao2020asgn}. Moreover, the application of GNNs accelerated the work in many related domains, including drug discovery \cite{wieder2020compact, feinberg2018potentialnet}, biology \cite{bove2020prediction}, physics \cite{shlomi2020graph}, and these GNNs reduced computational cost compared to traditional first-principles methods such as Density Functional Theory \cite{hao2020asgn, lu2019molecular}. In practice, there are many GNN variants that can be employed for molecular property prediction. Each variant is proposed based upon a distinct idea for feature learning of molecules. For example, GC (graph convolutional network) \cite{duvenaud2015convolutional} exploits neural architectures to generalise the chemical operation of the circular fingerprint to extract molecular features. In contrast, Weave \cite{kearnes2016molecular} and MPNN (message passing neural network) \cite{gilmer2017neural} have been proposed to learn molecular features by taking readout operations \cite{wu2020comprehensive} on atomic features. To learn atomic features, Weave applies global convolution operation, and MPNN uses the message passing process.

However, the hyperparameter selection is a direct impediment for GNNs to achieve excellent results. In general, the process of searching optimal hyperparameters is often a trial-and-error process. Traditionally, people used to adjust hyperparameters manually, but this requires domain experience and intuition. To extricate people from this predicament, random search (RS) has been employed for hyperparameter optimisation (HPO). In brief, RS draw hyperparameter values from uniform distributions within given ranges. The drawn hyperparameter values are evaluated on an objective function, and the one with the best performance will be selected when the given computational resource is exhausted. Although very simple, RS has been proved to be efficient for HPO of neural networks for many problems \cite{bergstra2011algorithms}. In recent years, there is an increasing number of strategies proposed for HPO. TPE \cite{bergstra2011algorithms} and CMA-ES \cite{hansen2016cma} are two state-of-the-art HPO algorithms, and they are proposed to improve the efficiency of search for promising hyperparameters by utilising the experience of previous trials. In this paper, a trial denotes the process of evaluating a hyperparameter setting on an objective function \cite{bergstra2012random}.

Research on HPO of GNNs for molecular property prediction is still in its infancy. For example, the pioneering work of GNN presented in \cite{duvenaud2015convolutional, gilmer2017neural, wu2018moleculenet} did not discuss the problem of HPO in detail. Meanwhile, most HPO methods have not been explored in terms of their efficiency on GNNs when facing this type of problems. Their performance may need to be further investigated because the sizes of molecular datasets vary from hundreds to thousands, which are far less than those of the datasets used for typical deep learning applications (e.g., image recognition, natural language processing). At the same time, predicting molecular properties requires more sophisticated neural architectures to process irregular molecular structures, which is different from image processing problems that have regular spatial patterns within image patches. Therefore, it has become necessary to explore the performance of existing HPO methods on GNNs in molecular domains. This motivates our research, and we conducted methodology comparison and experimental analysis for RS, TPE, and CMA-ES to assess their effects on GNNs as HPO methods. We expect that our research can inform researchers in both molecular machine learning and chemistry and material sciences. 

The contributions of our research are summarised as below:
\begin{itemize}
    \item We conducted systematic experiments to compare and analyse HPO methods including RS, TPE, and CMA-ES for GNN in the molecular domain in terms of their features, computational cost, and performance.
    \item Our research on HPO for GNN can be applied to a wider range of domains such as physical and social sciences.
    \item The outcomes of our research will contribute to the development of molecular machine learning \cite{pfluger2020molecular} as well as HPO for GNNs in general.

\end{itemize}

The rest of this paper is organized as follows. In Section \ref{sec: related work}, the related work of RS, TPE, and CMA-ES will be presented. In Section \ref{sec: theoretical analysis}, we will conduct a methodology comparison of RS, TPE, and CMA-ES. After that, the design of experiments and detailed experimental results will be described and discussed in Section \ref{sec: experiment investigation}. Finally, in Section \ref{sec: conclusion}, conclusions and future work will be given.

\section{Related Work}\label{sec: related work}

\subsection{Random Search}
\begin{algorithm}
$\operatorname{RS}\left(f, T,U\right)$ \tcp*[r]{\small $T$ the total number of trials, $U$ normal distribution}
 \For{$t \leftarrow 1$ \KwTo T}{
    $\lambda_{t} \leftarrow  U$\;
    
    $ \mathcal{L}_{t} \leftarrow \text {Evaluate } f(x_{train}, x_{valid}, \lambda_{t})$\;
    
    $\mathcal{H} \leftarrow \mathcal{H} \cup(\mathcal{L}_{t}, \lambda_{t})$\ \tcp*[r]{\small $\mathcal{H}$ historical optimization records}
    }
\Return{$\mathcal{H}$}
\caption{Random Search (RS) \cite{bergstra2012random}}
\label{alg:rs}
\end{algorithm}

Random Search (RS) \cite{bergstra2012random} is an approach that uses uniform probability in determining iterative procedures at the price of optimality \cite{zabinsky2010random}, and it is helpful for handling many ill‐structured global optimization problems with continuous and/or discrete variables \cite{zabinsky2010random} such as HPO. 

The process of applying RS for HPO is shown in Algorithm~\ref{alg:rs}. In \textbf{Line 3} of Algorithm \ref{alg:rs}, a solution $\lambda_{t}$ (i.e. a set of hyperparameter values) is sampled from a uniform distribution $U$, and then evaluated on objective function $f(x_{train}, x_{valid}, \lambda_{t})$ in \textbf{Line 4}, which is normally the most expensive step. The evaluation result $\mathcal{L}_{t}$ and the solution $\lambda_{t}$ is paired and recorded in $\mathcal{H}$. The procedures from \textbf{Line 3} to \textbf{Line 5} are iteratively executed $T$ times. Finally, the best solution is obtained by sorting historic solutions in $\mathcal{H}$ according to their corresponding $\mathcal{L}$ values.

Furthermore, Bergstra. J et al. \cite{bergstra2012random} holds the opinion that RS is the natural baseline for sequential HPO methods. Meanwhile, it is noted that Zelda B. \cite{zabinsky2010random} considered that RS is likely to be able to solve large-scale problems efficiently in a way that is not possible for deterministic algorithms. However, when using RS for HPO, the disadvantage is that its performance accompanied by high variance, and it may not produce satisfactory results given a larger search space and limited computational resources.

\subsection{TPE}

\begin{algorithm}
$\operatorname{TPE}\left(f, T, M_{0}, S\right)$ \tcp*[r]{\small $M$ surrogate model, $S$ sampling function}

 \For{$t \leftarrow 1$ \KwTo T}{
    $\lambda^{*}_{t} \leftarrow \operatorname{argmax}_{\lambda} S\left(\lambda, M_{t-1}\right)$\;
    $ \mathcal{L}_{t} \leftarrow \text {Evaluate } f(x_{train}, x_{valid}, \lambda^{*}_{t})$\;
    $\mathcal{H} \leftarrow \mathcal{H} \cup(\mathcal{L}_{t}, \lambda^{*}_{t}$)\;
    $\text { Fit a new model } M_{t} \text { to } \mathcal{H}$ \tcp*[r]{update $\mathcal{M}$}
    }
\Return{$\mathcal{H}$}
\caption{TPE \cite{bergstra2011algorithms}}
\label{alg:tpe}
\end{algorithm}
The problems of expensive evaluation on fitness function can be solved by sequential model-based global optimisation (SMBO) algorithms  \cite{bergstra2011algorithms, hutter2009automated, hutter2011sequential}. In HPO, a challenge is that the fitness function $f: \mathcal{X} \rightarrow \mathbb{R}$ may be expensive to evaluate given a trial of hyperparameters. By using model-based algorithms with a surrogate to approximate $f$ can reduce the evaluation cost. Typically, the core of an SMBO algorithm is to optimise the surrogate for the real fitness function, or some transformation of the surrogate. The two key components of SMBO algorithms are (1) what criterion is defined and optimized to obtain promising solutions given a model (or surrogate) of $f$, and (2) how $f$ can be approximated via historical trials/solutions. 

Tree-structured Parzen Estimator (TPE) \cite{bergstra2011algorithms} is an approach based on SMBO, as shown in Algorithm~\ref{alg:tpe}. Compared with RS, it makes the significant change in \textbf{Line 3}, in which solutions are sampled by $S$, instead of uniform distribution $U$. In $S$, many candidates $\lambda$ are drawn according to surrogate model $M$ and the one ($\lambda^{*}$) with the most promising performance evaluated by Expected Improvement (EI, introduce later) is returned \cite{jones2001taxonomy}. In \textbf{Line 4}, $\lambda^{*}$ is then evaluated on fitness function $f$ and recorded in $\mathcal{H}$ in \textbf{Line 5}. In \textbf{Line 6}, the surrogate $M$ is optimised to approximate the real fitness function by updated $\mathcal{H}$. Finally, the best solution can be obtained by sorting $\mathcal{H}$ after $T$ iterations. In the following paragraphs, we will review the most important work in TPE based on SMBO in detail.

In TPE, EI \cite{jones2001taxonomy} has been chosen as the criterion to guide the search for optimal solution(s), and it keeps the balance between exploitation and exploration during the search process. The utility function is defined as $u(\lambda)=\max \left(0, f^{\prime}-f(\lambda)\right)$, where $f^{\prime}$ denotes the output of the current best solution, and $\lambda$ is the solutions we want to find whose $f(\lambda)$ is expected as smaller as possible than $f^{\prime}$. The value of the difference between $f^{\prime}$ and $f(\lambda)$ will be return as a reward. In each iteration, the optimal solution $\lambda^{*}$ is given by $\mathrm{EI}_{y^{*}}(\lambda):=\int_{-\infty}^{\infty} \max \left(y^{*}-y, 0\right) p_{M}(y \mid \lambda) d y$, where $p_{M}(y \mid \lambda)$ is the surrogate of the real fitness function, and $y^{*}$ represents some quantile of the observed $y$ values.

Meanwhile, modelling of $p_{M}(y \mid \lambda)$ is costly, and TPE proposed an indirect way to model $p(\lambda \mid y)$ (Eq.~\ref{eq:1}). $p(\lambda \mid y)$ is defined by Eq.~\ref{eq:2}.where $\ell(\lambda)$ and $g(\lambda)$ are two density functions modelled by Parzen Estimator \cite{parzen1962estimation}, a non-parametric method to approximate the probability density function of a random variable. The collected observations are sorted by loss of $f$, and are divided into two groups based on some quantile. $\ell(\lambda)$ is generated by using the observations $\left\{\lambda^{(i)}\right\}$ such that the corresponding loss $f(\lambda)$ was less than $y^{*}$ and the remaining observations are used to generate $g(\lambda)$. In practice, a number of hyperparameter settings are sampled according to $\ell$, evaluated in term of $g(\lambda) / \ell(\lambda)$ , and the one that yields the minimum value under $\ell(\lambda)/g(\lambda)$ corresponding to the greatest EI is returned. This solution is then evaluated on the fitness function, and we call this process a trial. In this way, $g(\lambda)$ and $\ell(\lambda)$ are optimized according to the updated observation set, thus the exploration of optimal solutions moves to more promising regions of the whole search space by increasing the densities.

\begin{equation}\label{eq:1}
    p(y \mid \lambda)=\frac{p(\lambda \mid y) * p(y)}{p(\lambda)}
\end{equation}

\begin{equation}\label{eq:2}
    p(\lambda \mid y)=\left\{\begin{array}{ll}
\ell(\lambda) & \text { if } y<y^{*} \\
g(\lambda) & \text { if } y \geq y^{*}
\end{array}\right.
\end{equation}

In TPE, Tree-structure means that the hyperparameter space is tree-like, and the value chosen for one hyperparameter determines what hyperparameter will be chosen next and what values are available for it. 


\subsection{CMA-ES} 

\begin{algorithm}
$\operatorname{CMA-ES}(f, G, K, \mathcal{N})$ \tcp*[r]{\small $G$ the maximum number of generations, $K$ the size of population, $\mathcal{N}$ normal distribution}

 \For{$g \leftarrow 1$ \KwTo G}{
    \For{$k \leftarrow 1$ \KwTo K}{
        $\lambda^{g}_{k} \leftarrow \mathcal{N} $\;}
    \For{$k \leftarrow 1$ \KwTo K}{
    $\mathcal{L}_{k}^{g} \leftarrow \text {Evaluate } f(x_{train}, x_{valid}, \lambda^{g}_{k})$\;
    
        $\mathcal{H} \leftarrow \mathcal{H} \cup(\mathcal{L}^{g}_{k}, \lambda^{g}_{k})$\;
}
    $\text{Update } \mathcal{N}$
    }
\Return{$\mathcal{H}$}
\caption{CMA-ES \cite{hansen2016cma}}
\label{alg:cmaes}
\end{algorithm}
Covariance matrix adaptation evolution strategy (CMA-ES) \cite{hansen2016cma} is a derivative-free evolutionary algorithm for solving black-box optimization problems, and it has been applied for HPO with large-scale parallel GPU computational resources \cite{loshchilov2016cma, bergstra2011algorithms}. 

The pseudo-code of CMA-ES is shown in Algorithm~\ref{alg:cmaes}. In \textbf{Line 4}, a solution  $\boldsymbol{\lambda}_{k}^{(g)}$ is generated by sampling from a multivariate normal distribution $\mathcal{N}$ until the size of the population is satisfied, where $k$ denotes the index of offspring and $g$ for generation. Thereafter, in \textbf{Line 7}, the individuals of $\lambda^{g}$ will be evaluated on fitness function $f$ and recorded in $\mathcal{H}$. In \textbf{Line 10}, similar to TPE, it exploits the historical information $\mathcal{H}$ to optimise the search process. However, it is noted that CMA-ES optimises $\mathcal{N}$ rather than the surrogate, and we will discuss this in the following paragraphs.  

In CMA-ES, the multivariate distribution is re-defined. Specifically, a population of solutions $\lambda^{g+1}$ (i.e., individuals or offspring) is generated by sampling from a multivariate normal distribution $\mathcal{N}$ (Eq.~\ref{eq:multivar}). In Eq.~\ref{eq:multivar}, $\mathcal{N}\left(\mathbf{0}, \boldsymbol{C}^{(g)}\right)$ is a multivariate normal distribution with zero mean and covariance matrix $\boldsymbol{C}^{(g)}$. The latter decides the distribution shape and describes the correlations of the variables. Meanwhile, $\boldsymbol{m}^{(g)}$ represents the mean value which is the centroid of the distribution, and it determines the search region of the whole search space in generation $g$. $\sigma^{(g)}$ represents the step size which also decides the global variance; in other words, it controls the size of the region.

\begin{equation}\label{eq:multivar}
    \boldsymbol{x}_{k}^{(g+1)} \sim \boldsymbol{m}^{(g)}+\sigma^{(g)} \mathcal{N}\left(\mathbf{0}, \boldsymbol{C}^{(g)}\right) \quad \text { for } k=1, \ldots, K,
\end{equation}

To promote the efficiency of sampling, the key is to update $\boldsymbol{m}^{(g+1)}$, $C^{(g+1)}$, and $\sigma^{(g+1)}$ for the new generation. The mean is updated by the weighted average of $\mu$ selected individuals by $\boldsymbol{m}^{(g+1)}=\boldsymbol{m}^{(g)}+ \sum_{i=1}^{\mu} w_{i}\left(\boldsymbol{\lambda}_{i:K }^{(g+1)}-\boldsymbol{m}^{(g)}\right)$, where $w_{i}$ means corresponding weights to $\lambda_i$. The selection is according to the performance of individuals on the objective function. The novelty of CMA-ES is that it adapts the covariance matrix by combining rank-$\mu$-update and rank-one update \cite{akimoto2010bidirectional}. In this way, rank-µ update can efficiently make use of the information from the entire population. At the same time, rank-one update can be used to exploit the information of correlations among generations from the evolution path. This solution keeps the balance between less generations with a large population and more generation with a smaller population. Additionally, CMA-ES introduces the mechanism of step-size control based on evolution path (cumulative step-size adaptation of the global step-size) which aims to approximate the optimal overall step length efficiently by evolution path, because co-variance matrix may not be able to find the optimal overall step length efficiently.

Generally, CMA-ES imitates the biological evolution, assuming that no matter what kind of gene changes, the results (traits) always follow a Gaussian distribution of a variance and zero-mean. Meanwhile, the generated population is evaluated on a objective function. A part of well-performed individuals is selected to guide evolution, moving to the area where better individuals would be sampled with higher probability.

\section{Methodology Comparison}\label{sec: theoretical analysis}

\subsection{Common Features}
\begin{itemize}
    \item \textbf{Randomness} plays an important role in RS, TPE and CMA-ES. RS is supported by a number of independent uniform distributions with random sampling to explore hyperparameter space and find an optimal solution. TPE and CMA-ES both have exploitation and exploration mechanisms, which means they are given a more specific region of search space to explore compared with RS. TPE draws samples with randomness over the space of density function of $\ell(\lambda)$. Meanwhile, the sampling in CMA-ES is backed by a multivariate distribution.
    
    \item \textbf{Derivative-free} denotes that the approach does not use derivative information to guide the search for optimal solutions. For RS, TPE, and CMA-ES, as the last paragraph described, they search for the optimal solutions depending on drawing samples with randomness, rather than using gradient information as in the training of neural networks.
    
    \item \textbf{Termination Condition} As Section 2 shows, RS, TPE, and CMA-ES all have loops which means they need to preset the condition to stop the optimisation. However, this situation might cause a dilemma of balancing computational cost and performance.
\end{itemize}

\subsection{Specific Features}
\begin{itemize}
     \item \textbf{Uniform Distribution vs Gaussian Mixture Model vs Multivariate Normal Distribution } The uniform distribution is a symmetric probability distribution that gives a finite number of values with equal probability to be drawn. Meanwhile, in RS, the dimension of hyperparameter solutions corresponds to the required number of uniform distributions, and each uniform distribution is independent. In contrast, in CMA-ES, multivariate distribution is a distribution used to approximately describe a set of correlated real-valued random variables, each of which clusters around a mean value. Furthermore, TPE makes use of the Gaussian mixture model which assumes all the point are generated from a mixture of a number of Gaussian distributions with unknown parameters.
     
     \item \textbf{Model-based vs Model-free} These are two distinct approaches originally defined in reinforcement learning \cite{haith2013model}. RS is a very representative model-free approach that directly searches for better solutions via a process of trial-and-error. In contrast, TPE is a model-based approach that firstly uses density functions $\ell(x)$ and $g(x)$ to model hyperparameter space in terms of the surrogate, then searching solutions over the space of the functions. 
     
    \item \textbf{Bayesian Optimisation vs Evolutionary Strategy}
    The main idea of applying evolution strategies to black-box optimisation is to search through iterative adjustment of a multivariate normal distribution. The distribution is controlled by the mean and covariance, which are adjusted and moved to the area where better solutions could be sampled with higher probability. The adjustment generally has four main steps: sampling, evaluation, selecting good individuals, and updating the mean and co-variance by selected individuals. In contrast, in TPE, it starts by Bayesian optimisation to approximate the distribution of hyperparameters and objective function. Instead of using the Gaussian Process to model the distribution, TPE makes use of the Parzen estimator (i.e., kernel density estimation). The posterior distribution is unceasingly updated to approximate the real situation, and an acquisition function (TPE uses EI as the acquisition function) is used to approach the optimal solution.
\end{itemize}

\section{Experimental Investigation}\label{sec: experiment investigation}
In this section, we first describe the design of our systematic experiments. We then conducted four sets of experiments in Sections \ref{sec:computationalcost} and \ref{sec:performanceprioity}, to compare RS, TPE, and CMA-ES from the perspective of performance and computational cost. 

\subsection{Experimental Settings}
To investigate the performance of HPO methods for GNN on molecular property prediction, three representative datasets from Deep\-Chem \cite{Ramsundar-et-al-2019} (MoleculeNet)\cite{wu2018moleculenet} were selected in our experiments: ESOL (1128 records), FreeSolv (642 records), and Lipophilicity (4200 records), which respectively correspond to the tasks of predicting the following molecular properties: water solubility, hydration free energy, and octanol/water distribution coefficient. These properties are crucial in many problems. For example, in drug discovery, lipophilicity is an important property to reflect the affinity of a molecule, and it affects both membrane permeability and solubility \cite{lobo2020there}. Furthermore, the research presented in \cite{duong2012molecular} analyses molecular solubility data for exploring organic semiconducting materials. Therefore, the above three representative molecular property datasets are worth investigating and will benefit the research of many related problems. Using different sizes of datasets in experiments is helpful for us to conduct more comprehensive analyses. Meanwhile, there are many GNN variants, and we chose graph convolutional network (GC) \cite{duvenaud2015convolutional} because it was proposed considering the molecular domain background knowledge. The architectures of GC generalise the chemical operation of circular fingerprint \cite{glen2006circular} to extract molecular features. 


Four hyperparameters of GC are selected for HPO: batch size $s_b$, learning rate $l_r$,  the size of fully-connected layer $s_f$, and the size of graph convolution layer $s_g$. The selection is motivated by the related benchmark work presented in \cite{wu2018moleculenet} and considers molecular domain knowledge. By default setting, the number of graph convolution layers is two, but their sizes are the same, while the number of fully-connected layer is one. RS, TPE, and CMA-ES are implemented by Optuna \cite{akiba2019optuna}. The arguments of HPO methods in our experiments are set empirically or default values offered by Optuna. Meanwhile, considering that most practitioners/researchers do not have sufficient large-scale GPU computational resources as in industry (e.g., DeepMind, FAIR), we would like to assess the performance of HPO given limited resource, so our experiments all are conducted on a single GPU (GeForce GTX 1070), while the MedianPruner technique \cite{akiba2019optuna} is used to speed up HPO. We expect our experimental outcomes would inspire and help other people when they face similar HPO problems and are given limited computational resource.

In our experiments, every dataset is split into training, validation, and test sets with 80\%, 10\%, and 10\%. The training set is used to fit GC given a hyperparameter setting, and the validation set provides an unbiased evaluation of the hyperparameters during the search. The test set is used to evaluate the performance of HPO methods. The evaluation metric is the root mean square error (RMSE) of GC, and the loss function of GC defines the evaluation function. To make our experiments more persuasive, the best hyperparameter settings found by each method are given to GC, and then the GC is run for 30 times independently to calculate the mean of RMSEs on the training, validation, and test datasets. Meanwhile, to statistically analyse the difference between those means, we conducted corresponding $t$-tests, in which $t$ denotes $t$-value and $h$ represents the hypothesis. $t$-test in our all experiments is set with the significance level of $\alpha= 5\%$. When $h = 1$, the equal mean hypothesis is rejected, and a positive $t$ means the latter has better performance than the former; a negative $t$ represents that the former is better than the latter. In contrast, $h = 0$, the equal mean hypothesis is accepted. Empirically, during HPO, we found that the results of evaluating each trial often fluctuated, and to minimize this effect on HPO performance evaluation, we use the mean value of RMSEs from three repeated evaluations of GC as a single evaluation value.

\subsection{Computational Cost as a Primary Consideration}\label{sec:computationalcost}
In this section, we assess the performance of RS, CMA-ES, and TPE, while considering the computational cost as a priority. In HPO, any discussion and analysis of method performance without considering computational cost is argumentative because naive RS can find optimal solutions if given sufficient time and computational budget, and it is a highly friendly approach for parallel computing. Therefore, any comparison of HPO methods must be performed with acceptable computational cost. 

\subsubsection{General Experiments}
Different HPO methods employ different optimisation strategies. To compare them fairly, we first proposed assigning each of them with a total of 100 trials, assuming all HPO methods have equal computational cost. In other words, RS, TPE, and CMA-ES all have 100 opportunities to evaluate the hyperparameter settings on GC.

Tables ~\ref{tab:generalexp esol}$\sim$\ref{tab:generalexp lipo} summary the experiments of RS, TPE, and CMA-ES, respectively on three datasets: ESOL, FreeSolv, and Lipophilicity from DeepChem over the defined search space where batch size $s_b$ and the size of graph convolution layer $s_g$ range from 32 to 256 with the incremental step 32; learning rate $l_r$ is from 0.0001 to 0.0016 with the incremental step of 0.0001; the size of fully-connected layer $s_f$ is from 64 to 512, with the step of 64. This search space has $2^{13}$ solutions in total. Meanwhile, $t$-test with $\alpha=5\%$ is employed to determine if there is a significant difference between the means of RMSEs. $h=0$ denotes null hypothesis that the performance of GC under the two hyperparameter settings do not have significant difference; $h=1$ means rejecting the equal mean hypothesis. The $t$-test results of Tables ~\ref{tab:generalexp esol}$\sim$\ref{tab:generalexp lipo} are respectively summarised in Tables~\ref{tab:ttestesol}$\sim$\ref{tab:ttestlipo}.

In Tables~\ref{tab:generalexp esol} and~\ref{tab:generalexp freesolv}, TPE and CMA-ES did not present significant difference on performance on the test set (see Tables~\ref{tab:ttestesol} and~\ref{tab:ttestfreesolv}), but the hyperparameter settings they found have different $s_g$ and $s_f$ in both datasets. The negative value of $t$ means the former is better, for example, the row of RS-TPE in Table \ref{tab:ttestfreesolv}, the $t$ on the test set is -3.3167, which represents RS has a smaller RMSE value than TPE (i.e., RS is better than TPE); when the value of $t$ is positive, it means the latter is better than the former, for example, the row of RS-TPE in Table \ref{tab:ttestesol}, the $t$ on the test set is 1.8355, which means the performance of TPE is better than that of RS. A larger absolute value of $t$ indicates a bigger difference. Moreover, RS outperforms the other two methods in Table~\ref{tab:generalexp freesolv} with significant difference (see Table \ref{tab:ttestfreesolv}).  We consider the problem of FreeSolv may be more complex (the work research in \cite{asthagiri2003absolute} discussed the problem of deviations of calculating hydration free energy), and TPE and CMA-ES are constrained by the size of search space, the number of trials, and the size of the dataset, thus they got stuck in local optima. In contrast, RS uses a completely random strategy which is helpful to deal with this kind of special situation. However, we believe that CMA-ES and TPE would find better solutions if given more trials and a larger search space. 


In Table~\ref{tab:generalexp lipo}, TPE demonstrated better performance than CMA-ES and RS with significant difference (see Table~\ref{tab:ttestlipo}). The size of the Lipophilicity dataset is the largest in our experiments; compared with smaller datasets, the evaluation on validation set would return a result with less deviations, which is helpful for TPE and CMA-ES to improve and update their strategies for promising solutions. However, CMA-ES did not show excellent performance in all datasets in this group of experiments, and we considered that CMA-ES is based on the evolution strategy, which means it depends on unceasingly generating new offspring to find solutions, so 100 trials might restrict its performance. 

\begin{table}
\caption{The general experimental settings on the ESOL Dataset \\ \small Search space: $s_b$:32$\sim$256, $step$=32; $l_r$:0.0001$\sim$0.0016, $step$=0.0001; $s_g$:32$\sim$256, $step$=32; $s_f$:64$\sim$512, $step$=64}
\centering
\label{tab:generalexp esol}
\resizebox{\columnwidth}{!}{
\begin{tabular}{|l|l|l|l|l|l|} 
\hline
HPO Methods                        & Hyperparameters & Train                   & Validation              & Test                    &                             \\ 
\hline
\multirow{4}{*}{RS} & $s_g$=256       & \multirow{2}{*}{$\mathbf{0.2666}$} & \multirow{2}{*}{0.9067} & \multirow{2}{*}{0.8888} & \multirow{2}{*}{Mean RMSE}  \\ 
\cline{2-2}
                               & $s_f$=64        &                         &                         &                         &                             \\ 
\cline{2-6}
                               & $l_r$=0.0016    & \multirow{2}{*}{0.0364} & \multirow{2}{*}{0.0542} & \multirow{2}{*}{0.0411} & \multirow{2}{*}{Mean STD}   \\ 
\cline{2-2}
                               & $s_b$=64        &                         &                         &                         &                             \\ 
\hline
\multirow{4}{*}{TPE}           & $s_g$=192       & \multirow{2}{*}{0.3083} & \multirow{2}{*}{$\mathbf{0.8739}$} & \multirow{2}{*}{$\mathbf{0.8667}$} & \multirow{2}{*}{Mean RMSE}  \\ 
\cline{2-2}
                               & $s_f$=192       &                         &                         &                         &                             \\ 
\cline{2-6}
                               & $l_r$=0.0015    & \multirow{2}{*}{0.0534} & \multirow{2}{*}{0.0401} & \multirow{2}{*}{0.0476} & \multirow{2}{*}{Mean STD}   \\ 
\cline{2-2}
                               & $s_b$=32        &                         &                         &                         &                             \\ 
\hline
\multirow{4}{*}{CMA-ES}        & $s_g$=256       & \multirow{2}{*}{0.2939} & \multirow{2}{*}{$\mathbf{0.8739}$} & \multirow{2}{*}{0.8782} & \multirow{2}{*}{Mean RMSE}  \\ 
\cline{2-2}
                               & $s_f$=64        &                         &                         &                         &                             \\ 
\cline{2-6}
                               & $l_r$=0.0016    & \multirow{2}{*}{0.0458} & \multirow{2}{*}{0.0424} & \multirow{2}{*}{0.0562} & \multirow{2}{*}{Mean STD}   \\ 
\cline{2-2}
                               & $s_b$=32        &                         &                         &                         &                             \\
\hline
\end{tabular}}
\end{table}

\begin{table}
\centering
\caption{The general experimental settings on the FreeSolv Dataset \\ 
\small Search space: $s_b$:32$\sim$256, $step$=32; $l_r$:0.0001$\sim$0.0016, $step$=0.0001; $s_g$:32$\sim$256, $step$=32; $s_f$:64$\sim$512, $step$=64}
\label{tab:generalexp freesolv}
\resizebox{\columnwidth}{!}{
\begin{tabular}{|l|l|l|l|l|l|} 
\hline
HPO Methods                        & Hyperparameters & Train                   & Validation              & Test                    &                             \\ 
\hline
\multirow{4}{*}{RS} & $s_g$=256       & \multirow{2}{*}{0.6197} & \multirow{2}{*}{$\mathbf{1.2175}$} & \multirow{2}{*}{$\mathbf{1.1040}$} & \multirow{2}{*}{Mean RMSE}  \\ 
\cline{2-2}
                               & $s_f$=320       &                         &                         &                         &                             \\ 
\cline{2-6}
                               & $l_r$=0.0015    & \multirow{2}{*}{0.1248} & \multirow{2}{*}{0.1055} & \multirow{2}{*}{0.0995} & \multirow{2}{*}{Mean STD}   \\ 
\cline{2-2}
                               & $s_b$=32        &                         &                         &                         &                             \\ 
\hline
\multirow{4}{*}{TPE}           & $s_g$=160       & \multirow{2}{*}{0.6875} & \multirow{2}{*}{1.3425} & \multirow{2}{*}{1.2006} & \multirow{2}{*}{Mean RMSE}  \\ 
\cline{2-2}
                               & $s_f$=448       &                         &                         &                         &                             \\ 
\cline{2-6}
                               & $l_r$=0.0015    & \multirow{2}{*}{0.1854} & \multirow{2}{*}{0.1711} & \multirow{2}{*}{0.1212} & \multirow{2}{*}{Mean STD}   \\ 
\cline{2-2}
                               & $s_b$=32        &                         &                         &                         &                             \\ 
\hline
\multirow{4}{*}{CMA-ES}        & $s_g$=256       & \multirow{2}{*}{$\mathbf{0.5792}$} & \multirow{2}{*}{1.2721} & \multirow{2}{*}{1.1967} & \multirow{2}{*}{Mean RMSE}  \\ 
\cline{2-2}
                               & $s_f$=64        &                         &                         &                         &                             \\ 
\cline{2-6}
                               & $l_r$=0.0016    & \multirow{2}{*}{0.2653} & \multirow{2}{*}{0.1907} & \multirow{2}{*}{0.2128} & \multirow{2}{*}{Mean STD}   \\ 
\cline{2-2}
                               & $s_b$=32        &                         &                         &                         &                             \\
\hline
\end{tabular}}
\end{table}

\begin{table}
\centering
\caption{The general experimental settings on the Lipophilicity Dataset \\ \small Search space: $s_b$:32$\sim$256, $step$=32; $l_r$:0.0001$\sim$0.0016, $step$=0.0001; $s_g$:32$\sim$256, $step$=32; $s_f$:64$\sim$512, $step$=64}
\label{tab:generalexp lipo}
\resizebox{\columnwidth}{!}{
\begin{tabular}{|l|l|l|l|l|l|} 
\hline
HPO Methods                        & Hyperparameters & Train                   & Validation              & Test                    &                             \\ 
\hline
\multirow{4}{*}{RS} & $s_g$=96        & \multirow{2}{*}{0.2682} & \multirow{2}{*}{0.7024} & \multirow{2}{*}{0.6949} & \multirow{2}{*}{Mean RMSE}  \\ 
\cline{2-2}
                               & $s_f$=384       &                         &                         &                         &                             \\ 
\cline{2-6}
                               & $l_r$=0.001     & \multirow{2}{*}{0.0444} & \multirow{2}{*}{0.0279} & \multirow{2}{*}{0.0248} & \multirow{2}{*}{Mean STD}   \\ 
\cline{2-2}
                               & $s_b$= 64       &                         &                         &                         &                             \\ 
\hline
\multirow{4}{*}{TPE}           & $s_g$=224       & \multirow{2}{*}{$\mathbf{0.2475}$} & \multirow{2}{*}{$\mathbf{0.6914}$} & \multirow{2}{*}{$\mathbf{0.6655}$} & \multirow{2}{*}{Mean RMSE}  \\ 
\cline{2-2}
                               & $s_f$=192       &                         &                         &                         &                             \\ 
\cline{2-6}
                               & $l_r$=0.0015    & \multirow{2}{*}{0.0328} & \multirow{2}{*}{0.0229} & \multirow{2}{*}{0.0219} & \multirow{2}{*}{Mean STD}   \\ 
\cline{2-2}
                               & $s_b$=32        &                         &                         &                         &                             \\ 
\hline
\multirow{4}{*}{CMA-ES}        & $s_g$=32        & \multirow{2}{*}{0.3496} & \multirow{2}{*}{0.7191} & \multirow{2}{*}{0.7183} & \multirow{2}{*}{Mean RMSE}  \\ 
\cline{2-2}
                               & $s_f$=64        &                         &                         &                         &                             \\ 
\cline{2-6}
                               & $l_r$=0.0016    & \multirow{2}{*}{0.0425} & \multirow{2}{*}{0.0309} & \multirow{2}{*}{0.0245} & \multirow{2}{*}{Mean STD}   \\ 
\cline{2-2}
                               & $s_b$=32        &                         &                         &                         &                             \\
\hline
\end{tabular}}
\end{table}

\begin{table}
\centering
\caption{$t$-Test on the ESOL}
\label{tab:ttestesol}
\begin{tabular}{|c|c|c|c|c|c|c|} 
\hline
                                               & $t$                      & $h$ & $t$     & $h$              & $t$     & $h$              \\ 
\hline
HPO Methods & \multicolumn{2}{c|}{Train}     & \multicolumn{2}{c|}{Valid} & \multicolumn{2}{c|}{Test}  \\ 
\hline
RS - TPE                                       & -3.4671 & 1   & 2.6223  & 1                & 1.8355  & 1                \\ 
\hline
RS - CMA-ES                                    & -2.5080                  & 1   & 2.5666  & 1                & 0.7706  & 0                \\ 
\hline
TPE - CMA-ES                                   & 1.0984                   & 0   & -0.0007 & 0                & -0.8396 & 0                \\
\hline
\end{tabular}
\end{table}

\begin{table}
\centering
\caption{$t$-Test on the FreeSolv}
\label{tab:ttestfreesolv}
\begin{tabular}{|c|c|c|c|c|c|c|} 
\hline
                                               & $t$     & $h$              & $t$     & $h$              & $t$     & $h$              \\ 
\hline
HPO Methods & \multicolumn{2}{c|}{Train} & \multicolumn{2}{c|}{Valid} & \multicolumn{2}{c|}{Test}  \\ 
\hline
RS - TPE                                       & -1.6328 & 0                & -3.3492 & 1                & -3.3167 & 1                \\ 
\hline
RS - CMA-ES                                    & 0.7435  & 0                & -1.3487 & 0                & -2.1245 & 1                \\ 
\hline
TPE - CMA-ES                                   & 1.8011  & 1                & 1.4804  & 0                & 0.0855  & 0                \\
\hline
\end{tabular}
\end{table}

\begin{table}
\centering
\caption{$t$-Test on the Lipophilicity}
\label{tab:ttestlipo}
\begin{tabular}{|c|c|c|c|c|c|c|} 
\hline
             & $t$      & $h$             & $t$     & $h$              & $t$     & $h$              \\ 
\hline
HPO Methods  & \multicolumn{2}{c|}{Train} & \multicolumn{2}{c|}{Valid} & \multicolumn{2}{c|}{Test}  \\ 
\hline
RS - TPE     & 2.0146   & 0               & 1.6454  & 0                & 4.7676  & 1                \\ 
\hline
RS - CMA-ES  & -7.1251  & 1               & -2.1564 & 1                & -3.6088 & 1                \\ 
\hline
TPE - CMA-ES & -10.2330 & 1               & -3.8762 & 1                & -8.6244 & 1                \\
\hline
\end{tabular}
\end{table}

\subsubsection{Experiments on One Hour Runtime}
The same number of trials may not be able to assign the same computation cost for different HPO methods in practice, because different trials of hyperparameters may incur different computational cost on evaluation. For example, a larger value of $s_f$/$s_g$ means more trainable parameters, which will take more computational resource for the corresponding trial. Therefore, in this section, we design another set of experiments, in which we assign 1 hour time and the same hardware configuration to different HPO methods on the ESOL dataset with the same search space defined in Table 1, to see which method can find the best solution. 

Within one hour, the best trials of hyperparameters from the HPO methods were selected to configure GC, and this GC will be run 30 times, and the results and $t$-test are shown in Tables~\ref{tab:limitedtime} and \ref{tab:ttes1hour}. In Table~\ref{tab:limitedtime}, RS completed the largest number of trials, and the performance is approximately equal to the one shown in Table~\ref{tab:generalexp esol} because it completed almost 100 trials, which is similar to the previous experiment. We believe RS is efficient and stable in such a small search space. Furthermore, TPE obviously showed surprising performance with 54 trials to accomplish almost the same performance as shown in Table~\ref{tab:generalexp esol}. Additionally, it is noted that TPE found two different hyperparameter settings respectively shown in Table~\ref{tab:generalexp esol} and \ref{tab:limitedtime} but with almost the same performance on the test dataset. Meanwhile, as shown in Table \ref{tab:ttes1hour}, the performance of three HPO methods within 1 hour runtime has significant difference: TPE performs the best. CMA-ES did not reach our expectation that at least it should maintain the similar performance to RS, and we consider that CMA-ES might not be suitable for our particular HPO with insufficient computational cost and relatively small search space. Furthermore, it is noted the under-performing of CMA-ES may be alleviated by further exploring the "meta parameters" of CMA-ES, for example, the population size of CMA-ES. However, this seems to be an even more challenging "meta-HPO" problem, which is beyond the scope of this research. We will explore this in our future work.

\begin{table}
\centering
\caption{Experiments on the ESOL Dataset given one hour running time \\
\small Search Space: $s_b$:32$\sim$256, $step$=32; $l_r$:0.0001$\sim$0.0016, $step$=\\0.0001; $s_g$:32$\sim$256, $step$=32; $s_f$:64$\sim$512, $step$=64}
\label{tab:limitedtime}
\scalebox{0.624}{
\begin{tabular}{|l|l|l|l|l|l|l|} 
\hline
HPO Methods                       & Number of Trials    & Hyperparameters & Train                                    & Validation              & Test                                     &                             \\ 
\hline
\multirow{4}{*}{RS} & \multirow{4}{*}{96} & $s_g$=224          & \multirow{2}{*}{0.3301} & \multirow{2}{*}{0.8817} & \multirow{2}{*}{0.8994} & \multirow{2}{*}{Mean RMSE}  \\ 
\cline{3-3}
                              &                     & $s_f$=448       &                                          &                         &                                          &                             \\ 
\cline{3-7}
                              &                     & $l_r$=0.0008    & \multirow{2}{*}{0.0492} & \multirow{2}{*}{0.0457} & \multirow{2}{*}{0.0544}                  & \multirow{2}{*}{Mean STD}   \\ 
\cline{3-3}
                              &                     & $s_b$=32        &                                          &                         &                                          &                             \\ 
\hline
\multirow{4}{*}{TPE}          & \multirow{4}{*}{54} & $s_g$=256          & \multirow{2}{*}{$\mathbf{0.3193}$}                  & \multirow{2}{*}{$\mathbf{0.8605}$} & \multirow{2}{*}{$\mathbf{0.8634}$}                  & \multirow{2}{*}{Mean RMSE}  \\ 
\cline{3-3}
                              &                     & $s_f$=256       &                                          &                         &                                          &                             \\ 
\cline{3-7}
                              &                     & $l_r$=0.0014    & \multirow{2}{*}{0.0462}                  & \multirow{2}{*}{0.0400} & \multirow{2}{*}{0.0408}                  & \multirow{2}{*}{Mean STD}   \\ 
\cline{3-3}
                              &                     & $s_b$=32        &                                          &                         &                                          &                             \\ 
\hline
\multirow{4}{*}{CMA-ES}       & \multirow{4}{*}{63} & $s_g$=32           & \multirow{2}{*}{0.4287}                  & \multirow{2}{*}{0.9231} & \multirow{2}{*}{0.9688}                  & \multirow{2}{*}{Mean RMSE}  \\ 
\cline{3-3}
                              &                     & $s_f$=512       &                                          &                         &                                          &                             \\ 
\cline{3-7}
                              &                     & $l_r$=0.0016    & \multirow{2}{*}{0.0933}                  & \multirow{2}{*}{0.0706} & \multirow{2}{*}{0.0845}                  & \multirow{2}{*}{Mean STD}   \\ 
\cline{3-3}
                              &                     & $s_b$=32        &                                          &                         &                                          &                             \\
\hline
\end{tabular}}
\end{table}

\begin{table}
\centering
\caption{$t$-Test on experiments on the ESOL Dataset given one hour running time}
\label{tab:ttes1hour}
\begin{tabular}{|c|c|c|c|c|c|c|} 
\hline
             & $t$     & $h$              & $t$     & $h$              & $t$     & $h$              \\ 
\hline
HPO Methods  & \multicolumn{2}{c|}{Train} & \multicolumn{2}{c|}{Valid} & \multicolumn{2}{c|}{Test}  \\ 
\hline
RS - TPE     & 0.8573  & 0                & 1.8774  & 1                & 2.8449  & 1                \\ 
\hline
RS - CMA-ES  & -5.0338 & 1                & -2.6530 & 1                & -3.7178 & 1                \\ 
\hline
TPE - CMA-ES & -5.6546 & 0                & -4.1563 & 1                & -6.0439 & 1                \\
\hline
\end{tabular}
\end{table}

\subsection{Performance as Primary Consideration}\label{sec:performanceprioity}
In this section, we design another group of experiments to explore RS, TPE and CMA-ES with performance as a primary consideration by providing as much as possible computational cost.

\subsubsection{Experiments on Repeated HPO Runs}\label{sec:exponrepeatedhporuns}
In order to purely compare HPO performance, we designed to respectively run three HPO methods on the ESOL dataset for 10 times independently, and each time is assigned with 100 trials and keep the search space the same as that defined in Table 1. The performance is evaluated by calculating the mean of RMSE values of the best trial from every 100 trials. We did not run it on the test data set because the 10 times RMSEs correspond to 10 different hyperparameter settings. Our purpose is to discover the capability of those methods to fit HPO problem, and the results and t-test are shown in Tables~\ref{tab:performance} and \ref{tab:tteshporepeat}. TPE again outperforms RS and CMA-ES, and it also presents more stable performance as with less standard deviation. $t$-test in Table~\ref{tab:tteshporepeat} shows the same outcome on the test dataset to Table~\ref{tab:generalexp esol} that RS and CMA-ES in this problem and search space have similar performance (no significant difference). So, we believe the CMA-ES still has room for improvement for its performance in our experiments.

\begin{table}
\centering
\caption{Experiments on the ESOL Dataset with performance as primary consideration}
\label{tab:performance}
\begin{tabular}{|l|l|l|l|} 
\hline
             & RS     & TPE    & CMA-ES  \\ 
\hline
Mean RMSE & 0.8529 & $\mathbf{0.8190}$ & 0.8469  \\ 
\hline
Std          & 0.0169 & 0.0090 & 0.0169  \\
\hline
\end{tabular}
\end{table}

\begin{table}
\centering
\caption{$t$-Test on ESOL with performance as primary consideratoin}
\label{tab:tteshporepeat}
\begin{tabular}{|c|c|c|} 
\hline
HPO Methods  & $t$     & $h$  \\ 
\hline
RS - TPE     & 5.2891  & 1    \\ 
\hline
RS - CMA-ES  & 0.7464  & 0    \\ 
\hline
TPE - CMA-ES & -4.3625 & 1    \\
\hline
\end{tabular}
\end{table}

\subsubsection{Experiments on A Larger Search Space}
To further investigate the performance of the HPO methods, we increased the search space so that $s_b$ and $s_g$ range from 8 to 512 with the step size of 8; $l_r$ is changed to from 0.0001 to 0.0032 with the step size of 0.0001; $s_f$ is increased from 32 to 1024 with the step size of 32. The new search space has $2^{22}$ configurations, and the increments become small and the value ranges are increased. The experimental details are shown in Tables~\ref{tab:searchspaceesol}, \ref{tab:searchspacefreesolv}, and \ref{tab:searchspacelipo}; while the corresponding $t$-test are presented in Tables \ref{tab:ttestesollarger}, \ref{tab:ttestfreesolvlarger}, and \ref{tab:ttestlipolarger}.

In the three datasets, in general, the RMSEs for RS, TPE, and CMA-ES on the test datasets have improved compared with the experiments in Section \ref{sec:computationalcost} given the same number of trials. Meanwhile, by observing the results on the validation and test datasets for all three datasets, we do not see over-fitting issues.  In ESOL,  TPE and CMA-ES have almost the same performance, and both of them are better than RS as indicated by $t$-test (see Table~\ref{tab:ttestesollarger}). In addition, in FreeSolv, three HPO methods show no significant difference performance, while TPE and CMA-ES made the improvements compared with the previous experiments (see Table~\ref{tab:generalexp freesolv}). It is sensible that a potentially complex problem should be given a large search space to find the most suitable hyperparameters. In Table \ref{tab:searchspacelipo}, the rank of performance of the methods is still that TPE is the best, and RS is better than CMA-ES.

\begin{table}
\centering
\caption{Experiments in larger search space on the ESOL Dataset \\
\small $s_b$:8$\sim$512, $step$=8; $l_r$:0.0001$\sim$0.0032, $step$=0.0001; $s_g$:8$\sim$512, $step$=8; $s_f$:32$\sim$1024, $step$=32}
\resizebox{\columnwidth}{!}{
\begin{tabular}{|l|l|l|l|l|l|} 
\hline
HPO Methods                        & Hyperparameters & Train                   & Validation              & Test                    &                        \\ 
\hline
\multirow{4}{*}{RS} & $s_g$=384                        & \multirow{2}{*}{$\mathbf{0.3190}$} & \multirow{2}{*}{0.8727} & \multirow{2}{*}{0.8479} & \multirow{2}{*}{Mean RMSE}  \\ 
\cline{2-2}
                               & $s_f$=160                        &                         &                         &                         &                             \\ 
\cline{2-6}
                               & $l_r$=0.0016                     & \multirow{2}{*}{0.0323} & \multirow{2}{*}{0.0310} & \multirow{2}{*}{0.0453} & \multirow{2}{*}{Mean STD}   \\ 
\cline{2-2}
                               & $s_b$=24                         &                         &                         &                         &                             \\ 
\hline
\multirow{4}{*}{TPE}           & $s_g$=312                        & \multirow{2}{*}{0.5089} & \multirow{2}{*}{$\mathbf{0.8203}$} & \multirow{2}{*}{0.781}  & \multirow{2}{*}{Mean RMSE}  \\ 
\cline{2-2}
                               & $s_f$=224                        &                         &                         &                         &                             \\ 
\cline{2-6}
                               & $l_r$=0.003                      & \multirow{2}{*}{0.1281} & \multirow{2}{*}{0.096}  & \multirow{2}{*}{0.057}  & \multirow{2}{*}{Mean STD}   \\ 
\cline{2-2}
                               & $s_b$=8                          &                         &                         &                         &                             \\ 
\hline
\multirow{4}{*}{CMA-ES}        & $s_g$=512                        & \multirow{2}{*}{0.5793} & \multirow{2}{*}{0.848}  & \multirow{2}{*}{$\mathbf{0.7772}$} & \multirow{2}{*}{Mean RMSE}  \\ 
\cline{2-2}
                               & $s_f$=1024                       &                         &                         &                         &                             \\ 
\cline{2-6}
                               & $l_r$=0.0032                     & \multirow{2}{*}{0.1529} & \multirow{2}{*}{0.1247} & \multirow{2}{*}{0.097}  & \multirow{2}{*}{Mean STD}   \\ 
\cline{2-2}
                               & $s_b$=8                          &                         &                         &                         &                             \\
\hline
\end{tabular}}
\label{tab:searchspaceesol}
\end{table}

\begin{table}
\centering
\caption{Experiments in larger search space on the FreeSolv Dataset \\
\small $s_b$:8$\sim$512, $step$=8; $l_r$:0.0001$\sim$0.0032, $step$=0.0001; $s_g$:8$\sim$512, $step$=8; $s_f$:32$\sim$1024, $step$=32}
\label{tab:searchspacefreesolv}
\resizebox{\columnwidth}{!}{
\begin{tabular}{|l|l|l|l|l|l|} 
\hline
HPO Methods                        & Hyperparameters & Train                    & Validation              & Test                    &                               \\ 
\hline
\multirow{4}{*}{RS} & $s_g$=200       & \multirow{2}{*}{$\mathbf{0.3747}$}  & \multirow{2}{*}{1.2412} & \multirow{2}{*}{1.0880} & \multirow{2}{*}{Mean RMSE}    \\ 
\cline{2-2}
                               & $s_f$=64        &                          &                         &                         &                               \\ 
\cline{2-6}
                               & $l_r$=0.0030    & \multirow{2}{*}{0.0684}  & \multirow{2}{*}{0.1152} & \multirow{2}{*}{0.0990} & \multirow{2}{*}{Mean STD}     \\ 
\cline{2-2}
                               & $s_b$=48        &                          &                         &                         &                               \\ 
\hline
\multirow{4}{*}{TPE}           & $s_g$=424       & \multirow{2}{*}{0.6144}  & \multirow{2}{*}{$\mathbf{1.1288}$} & \multirow{2}{*}{$\mathbf{1.0620}$} & \multirow{2}{*}{Mean RMSE~~}  \\ 
\cline{2-2}
                               & $s_f$=224       &                          &                         &                         &                               \\ 
\cline{2-6}
                               & $l_r$=0.0008    & \multirow{2}{*}{0.0951}  & \multirow{2}{*}{0.1163} & \multirow{2}{*}{0.1115} & \multirow{2}{*}{Mean STD}     \\ 
\cline{2-2}
                               & $s_b$=16        &                          &                         &                         &                               \\ 
\hline
\multirow{4}{*}{CMA-ES}        & $s_g$=512       & \multirow{2}{*}{0.6973}  & \multirow{2}{*}{1.2329} & \multirow{2}{*}{1.0835} & \multirow{2}{*}{Mean RMSE~~}  \\ 
\cline{2-2}
                               & $s_f$=32        &                          &                         &                         &                               \\ 
\cline{2-6}
                               & $l_r$=0.0032    & \multirow{2}{*}{0.07819} & \multirow{2}{*}{0.1306} & \multirow{2}{*}{0.1073} & \multirow{2}{*}{Mean STD}     \\ 
\cline{2-2}
                               & $s_b$=8         &                          &                         &                         &                               \\
\hline
\end{tabular}}
\end{table}

\begin{table}
\centering
\caption{Experiments in larger search space on the Lipophilicity Dataset\\
\small $s_b$:8$\sim$512, $step$=8; $l_r$:0.0001$\sim$0.0032, $step$=0.0001; $s_g$:8$\sim$512, $step$=8; $s_f$:32$\sim$1024, $step$=32}
\label{tab:searchspacelipo}
\resizebox{\columnwidth}{!}{
\begin{tabular}{|l|l|l|l|l|l|} 
\hline
HPO Methods                        & Hyperparameters & Train                                    & Validation              & Test                    &                             \\ 
\hline
\multirow{4}{*}{RS} & $s_g$=312       & \multirow{2}{*}{0.2570}                  & \multirow{2}{*}{$\mathbf{0.6736}$} & \multirow{2}{*}{0.6552} & \multirow{2}{*}{Mean RMSE}  \\ 
\cline{2-2}
                               & $s_f$=32        &                                          &                         &                         &                             \\ 
\cline{2-6}
                               & $l_r$=0.0031    & \multirow{2}{*}{0.0240}                  & \multirow{2}{*}{0.0285} & \multirow{2}{*}{0.0223} & \multirow{2}{*}{Mean STD}   \\ 
\cline{2-2}
                               & $s_b$=32        &                                          &                         &                         &                             \\ 
\hline
\multirow{4}{*}{TPE}           & $s_g$=496       & \multirow{2}{*}{$\mathbf{0.2413}$}                  & \multirow{2}{*}{0.6786} & \multirow{2}{*}{$\mathbf{0.6395}$} & \multirow{2}{*}{Mean RMSE}  \\ 
\cline{2-2}
                               & $s_f$=32        &                                          &                         &                         &                             \\ 
\cline{2-6}
                               & $l_r$=0.0022    & \multirow{2}{*}{0.0188}                  & \multirow{2}{*}{0.0195} & \multirow{2}{*}{0.0193} & \multirow{2}{*}{Mean STD}   \\ 
\cline{2-2}
                               & $s_b$=24        &                                          &                         &                         &                             \\ 
\hline
\multirow{4}{*}{CMA-ES}        & $s_g$=248       & \multirow{2}{*}{0.2442}                  & \multirow{2}{*}{0.6931} & \multirow{2}{*}{0.6826} & \multirow{2}{*}{Mean RMSE}  \\ 
\cline{2-2}
                               & $s_f$=480       &                                          &                         &                         &                             \\ 
\cline{2-6}
                               & $l_r$=0.0015    & \multirow{2}{*}{0.0430} & \multirow{2}{*}{0.0194} & \multirow{2}{*}{0.0167} & \multirow{2}{*}{Mean STD}   \\ 
\cline{2-2}
                               & $s_b$=120       &                                          &                         &                         &                             \\
\hline
\end{tabular}}
\end{table}

\begin{table}
\centering
\caption{$t$-test on the ESOL in larger search space}
\label{tab:ttestesollarger}
\begin{tabular}{|c|c|c|c|c|c|c|} 
\hline
             & $t$      & $h$             & $t$     & $h$              & $t$    & $h$               \\ 
\hline
HPO Methods  & \multicolumn{2}{c|}{Train} & \multicolumn{2}{c|}{Valid} & \multicolumn{2}{c|}{Test}  \\ 
\hline
RS - TPE     & -7.7384  & 1               & 2.7832  & 1                & 4.9451 & 1                 \\ 
\hline
RS - CMA-ES  & -10.7815 & 1               & 1.0330  & 0                & 3.5578 & 1                 \\ 
\hline
TPE - CMA-ES & -2.1105  & 1               & -0.9456 & 0                & 0.1855 & 0                 \\
\hline
\end{tabular}
\end{table}

\begin{table}
\centering
\caption{$t$-test on the FreeSolv in larger search space}
\label{tab:ttestfreesolvlarger}
\begin{tabular}{|c|c|c|c|c|c|c|} 
\hline
             & $t$      & $h$             & $t$     & $h$              & $t$     & $h$              \\ 
\hline
HPO Methods  & \multicolumn{2}{c|}{Train} & \multicolumn{2}{c|}{Valid} & \multicolumn{2}{c|}{Test}  \\ 
\hline
RS - TPE     & -11.0149 & 1               & 3.6976  & 1                & 0.9385  & 0                \\ 
\hline
RS - CMA-ES  & -16.7183 & 1               & 0.2580  & 0                & 0.1661  & 0                \\ 
\hline
TPE - CMA-ES & -3.6269  & 1               & -3.2042 & 1                & -0.7475 & 0                \\
\hline
\end{tabular}
\end{table}

\begin{table}
\centering
\caption{$t$-test on the Lipophilicity in larger search space}
\label{tab:ttestlipolarger}
\begin{tabular}{|c|c|c|c|c|c|c|} 
\hline
             & $t $      & $h$              & $t $      & $h$              & t       & $h$              \\ 
\hline
HPO Methods  & \multicolumn{2}{c|}{Train} & \multicolumn{2}{c|}{Valid} & \multicolumn{2}{c|}{Test}  \\ 
\hline
RS - TPE     & 2.7588  & 1                & -0.7815 & 0                & 2.7989  & 1                \\ 
\hline
RS - CMA-ES  & 1.3964  & 0                & 3.0423  & 1                & -5.2825 & 1                \\ 
\hline
TPE - CMA-ES & -0.3299 & 0                & -2.8299 & 1                & -8.9802 & 1                \\
\hline
\end{tabular}
\end{table}

\section{Conclusion and Future Work}\label{sec: conclusion}
Overall, our experimental results indicate that TPE is the most suited HPO method for GNN as applied to our molecular property prediction problems given limited computational resources. Meanwhile, RS is the simplest method but can achieve comparable performance against TPE and CMA-ES. In our future work, facing molecular problems on small datasets, the use of CMA-ES also deserves further investigation, and we believe that CMA-ES, RS, and TPE will have very similar performance given more computational budget. Furthermore, as mentioned in Section \ref{sec:exponrepeatedhporuns}, the selection of the "meta-parameters" for HPO methods deserve more research; we will investigate the impact of HPO methods' meta-parameter values on their performance.

Finally, we expect that our work will help people from various fields (e.g., machine learning, chemistry, materials science) when they are facing similar type interdisciplinary problems. As the applications of GNNs have been explored in many areas and indeed benefited the research in those areas, we believe that our research outcomes would give them valuable insights to facilitate their research.

\section*{ACKNOWLEDGMENTS}
This research is supported by the Engineering and Physical Sciences Research Council (EPSRC) funded Project on New Industrial Systems: Manufacturing Immortality (EP/R020957/1). The authors are also grateful to the Manufacturing Immortality consortium.

\section*{Data Statement}
All data used in our experiments are from MoleculeNet \cite{wu2018moleculenet}, which are publicly available in \url{http://moleculenet.ai/datasets-1}.